\documentclass[english,prd,twocolumn,superscriptaddress,floatfix,nofootinbib,preprintnumbers]{revtex4}
\usepackage{graphicx}
\usepackage{amsmath,amsthm,amssymb}
\usepackage{bm}

\usepackage{amsfonts}
\usepackage{dcolumn}
\usepackage{hyperref}

\def\be{\begin{equation}}
\def\ee{\end{equation}}
\def\ba{\begin{eqnarray}}
\def\ea{\end{eqnarray}}
\def\bs{\begin{subequations}}
\def\es{\end{subequations}}

\def\cy{{\cal Y}}

\newcommand{\pa}{{\partial}}

\usepackage{color}

\def\car{{\cal R}}
\def\cy{{\cal Y}}

\makeatother

\usepackage{babel}
\makeatother

\begin{document}

{\hbox to\hsize{
\vbox{\noindent IPMU12-0089}}}
\vglue.2in

\title{Consistency of inflation and preheating in $F({\cal R})$ supergravity}

\author{Sergei V.~Ketov}

\affiliation{Department of Physics, Tokyo Metropolitan University, Minami-ohsawa 1-1, 
Hachioji-shi, Tokyo 192-0397, Japan}

\affiliation{Kavli Institute for the Physics and Mathematics of the Universe (IPMU), 
The University of Tokyo, Kashiwanoha 5-1-5, Kashiwa-shi, Chiba 277-8568, Japan}

\author{Shinji Tsujikawa}

\affiliation{Department of Physics, Faculty of Science,
Tokyo University of Science,
1-3, Kagurazaka, Shinjuku-ku, Tokyo 162-8601, Japan}

\begin{abstract}

We study inflation and preheating in $F({\cal R})$ supergravity 
characterized by two mass scales of a scalar degree of 
freedom (scalaron): $M$ (associated with the inflationary era) and 
$m$ (associated with the preheating era).
The allowed values of the masses $M$ and $m$ are derived
from the amplitude of the CMB temperature anisotropies.
We show that our model is consistent with the joint 
observational constraints of WMAP and other measurements 
in the regime where a sufficient amount of inflation 
(with the number of e-foldings larger than 50) is realized.
In the low-energy regime relevant to preheating, 
we derive the effective scalar potential 
in the presence of a pseudo-scalar field $\chi$ coupled 
to the inflaton (scalaron) field $\phi$.
If $m$ is much larger than $M$, we find that there exists
the preheating stage in which the field perturbations 
$\delta \chi$ and $\delta \phi$ rapidly grow by a broad
parametric resonance.

\end{abstract}

\date{\today}


\maketitle

\section{Introduction}

The gravitational theory dubbed $f(R)$ gravity, where the 
Lagrangian $f$ is a function of the Ricci scalar $R$, is the viable 
phenomenological framework to describe inflation and reheating 
in the early Universe \cite{Starobinsky80} 
(see also Refs.~\cite{RR69,Kerner} for early works). 
For the simple model described by the Lagrangian $f(R)=R-R^2/(6M^2)$, 
where $M$ corresponds to the mass of a scalar degree of freedom
(``scalaron''), the presence of the quadratic term $R^2/(6M^2)$ leads 
to cosmic acceleration followed by a gravitational 
reheating \cite{S82}.
Moreover the predicted power spectra of scalar and tensor 
perturbations in this model are compatible with the observations of 
the CMB temperature anisotropies \cite{spectra}
(see Refs.~\cite{Soti,DTreview,Clifton} for reviews).

An embedding (or derivation) of a viable $f(R)$ gravity model from 
a more fundamental physical theory naturally leads to 
supergravity (the theory of local supersymmetry)
as the first step, because supersymmetry is the leading and well
motivated proposal for new physics beyond the 
Standard Model of elementary particles.
Supergravity is also the low-energy effective field theory 
of superstrings.

A manifestly ${\cal N}=1$ locally supersymmetric (minimal) extension of $f(R)$ 
gravity was recently constructed in a curved superspace \cite{Gates}.
This was dubbed $F({\cal R})$ supergravity, where 
$F({\cal R})$ is a holomorphic function of the covariantly-chiral
scalar curvature superfield ${\cal R}$.
The basic features of $F(\car)$ supergravity and some of its non-trivial 
models were systematically studied 
in Refs.~\cite{Ketov10,Watanabe,Kreview,Ketov12}.

The first phenomenologically viable inflationary model based on $F(\car)$ 
supergravity was proposed in Ref.~\cite{Ketov11}. 
In the high-energy regime relevant to inflation this is similar to the 
Starobinsky's $f(R)$ model with the quadratic term $-R^2/(6M^2)$ \cite{Starobinsky80}, 
but the correction of the form $(-R)^{3/2}$ is also present.
The viability of this model was proven 
in certain limit of its parameter space, but phenomenological and 
observational bounds on the parameter values were not found yet. 

In the low-energy regime relevant to reheating the proposed model
of $F(\car)$ supergravity \cite{Ketov11} is approximately described by 
the Lagrangian $f(R)=R-R^2/(6m^2)$, where the mass
scale $m$ is not identical to the mass $M$ in the high-energy regime.
In this case the dynamics of reheating should be different from that 
studied in Refs.~\cite{S82} for the original Starobinsky's $f(R)$ model.
It was conjectured in Ref.~\cite{Ketov11} that the model \cite{Ketov11}
gives rise to efficient preheating after inflation, 
but it was not proven or verified by concrete calculations.

In this paper we extend the analysis of Ref.~\cite{Ketov11} by providing physical 
interpretation to the parameters of the model \cite{Ketov11}, confirm its consistency 
and viability, and find observational bounds on its parameters. 
In the low-curvature regime we also derive the effective scalar potential 
of a scalaron field $\phi$ coupled to a pseudo-scalar field $\chi$.
This potential is employed for numerical analysis of the preheating stage after inflation.  
We show the existence of a broad parametric resonance, by which both the field 
perturbations $\delta \chi$ and $\delta \phi$ are amplified in the parameter 
region where $m$ is much larger than $M$.

Our paper is organized as follows. 
In Sec.~\ref{modelsec} we describe how $f(R)$ gravity arises from
$F(\car)$ supergravity and review the model \cite{Ketov11} by adding more details. 
In Sec.~\ref{infsec} we study the inflationary dynamics in the model \cite{Ketov11} 
under the slow-roll approximation and provide phenomenological bounds 
on the model parameters. 
In Sec.~\ref{obsersec} our model \cite{Ketov11} is confronted with 
the observational tests of CMB and other measurements.  
In Sec.~\ref{scalarpotensec} we compute the effective scalar potential 
of two physical scalars in the low-energy regime.
In Sec.~\ref{reheatingsec} we numerically study the dynamics of 
preheating of the multi-field system.
Our conclusion is given in Sec.~\ref{concludesec}.

\section{Model}
\label{modelsec}

The action of $F({\cal R})$ supergravity in the chiral (curved) 
${\cal N}=1$ superspace of $(1+3)$-dimensional spacetime 
is given by \cite{Gates} 
\be
S=\int d^4 x\,d^2 \theta\, {\cal E}\,
F({\cal R})+{\rm H.c.}\,,
\label{frsup}
\ee
where $F({\cal R})$ is a holomorphic function of the covariantly-chiral
scalar curvature superfield ${\cal R}$, and ${\cal E}$ is the chiral 
superspace density \cite{SGbook}. The scalar curvature $R$ appears as the field coefficient of the $\theta^2$ term in the superfield ${\cal R}$.
We use the metric signature $(+, -, -, -)$, so that the sign of $R$ is opposite 
to that of Ref.~\cite{DTreview}. See Ref.~\cite{Kreview} for more details about 
our notation and $F({\cal R})$ supergravity. 

The chiral superspace density (in a Wess-Zumino type gauge) reads
\be
{\cal E}=e(x) \left[ 1-2i \sigma_a \bar{\psi}^{a} (x)
+\theta^2 B(x) \right]\,,
\ee
where $e=\sqrt{-{\rm det}\,g_{\mu \nu}}$, $g_{\mu \nu}$ is 
a spacetime metric, $\psi^{a}_{\alpha}=e^{a}_{\mu} \psi^{\mu}_{\alpha}$
is a chiral gravitino, and $B=S-iP$ is the complex scalar 
auxiliary field. When dropping the contribution of gravitinos ($\psi_{\mu}=0$),
there is the following formula for the superfield Lagrangian ${\cal L} (x, \theta)$:
\be
S=\int d^4 x\,d^2 \theta\, {\cal E}
{\cal L}=\int d^4 x\, e \left[ B {\cal L}_1 (x)
+{\cal L}_2 (x) \right]\,,
\label{sfor}
\ee
where
\be
{\cal L}_1 (x)={\cal L}|\,,\qquad
{\cal L}_2 (x)=\nabla^2 {\cal L}|\,.
\ee
Here the vertical bars denote the leading field components 
of superfields. In particular one has 
\ba
{\cal R}| &=& \frac{\bar{B}}{3M_{\rm pl}}=
\frac{1}{3M_{\rm pl}} (S+iP)\,,\\
\nabla^2 {\cal R}| &=&
\frac13 R+\frac{4\bar{B}B}{9M_{\rm pl}^2}\,,
\ea
where $M_{\rm pl}=2.435 \times 10^{18}$~GeV 
is the reduced Planck mass.

Applying the formula (\ref{sfor}) to the action (\ref{frsup}), 
we obtain the bosonic part of the supersymmetric Lagrangian of
Eq.~(\ref{frsup}) in the form \cite{Ketov10}
\be
L=3X F(\bar{X})+\left( \frac13 R +4\bar{X}X \right)
F'(\bar{X}) +{\rm H.c.}\,,
\label{boso}
\ee
where $X \equiv B/(3M_{\rm pl})$.
Here and in the next Secs.~\ref{infsec}
and \ref{obsersec} we focus on the further reduced case 
in which the auxiliary field $B$ is real, i.e. $X=\bar{X}$, by ignoring 
the complex nature of the supergravity superfields. In physical terms, 
it amounts to dropping the bosonic degree of freedom which is the pseudo-scalar 
superpartner of scalaron (see Sec.~\ref{scalarpotensec}), 
in addition to dropping all fermionic degrees 
of freedom. Then the Lagrangian (\ref{boso}) reduces to 
\be
L=6X F(X)+2 \left( \frac13 R+4X^2 \right) F'(X)\,.
\label{boso2}
\ee
Variation of the action (\ref{boso2}) with respect to $X$
gives rise to the following {\it algebraic} relation between $X$ and $R$:
\be
3F(X)+11X F'(X)+\left( \frac13 R+4X^2 \right) F''(X)=0\,.
\label{Feq}
\ee

It is natural to expand the function $F({\cal R})$ into power
series of ${\cal R}$, i.e. 
$F({\cal R})=\sum_{i=0}^{n}c_i{\cal R}^{i}$, with some (generically complex)
coefficients $c_i$. Since we are not interested in the parity-violating terms
in this paper, we assume that all the coefficients $c_i$ are real.
For example, the choice $F({\cal R})=f_0-f_1 {\cal R}/2$ gives rise to 
the standard ${\cal N}=1$ supergravity with a negative cosmological constant. 
Slow-roll inflation can be realized by the following function \cite{Ketov11}:
\be
F({\cal R})=-\frac12 f_1 {\cal R}+\frac12 f_2 {\cal R}^2
-\frac16 f_3 {\cal R}^3\,,
\label{Flag}
\ee
where $f_{1,2,3}$ are positive constants having dimensions
of [mass]$^2$, [mass]$^1$, and [mass]$^0$ respectively.  
In this case the Lagrangian (\ref{boso2}) yields
\ba
L &=& -\frac13 f_1 R+\frac23 f_2 RX-\left( 7f_1+\frac13 f_3 R 
\right) X^2 \nonumber \\
& &+11f_2 X^3-5f_3 X^4\,.
\label{bosof}
\ea
It follows from Eq.~(\ref{Feq}) that 
\be
X^3-\frac{33f_2}{20f_3}X^2+\frac{1}{30} (R+R_0)X
-\frac{f_2}{30 f_3}R=0\,,
\label{Xeq}
\ee
where 
\be \label{rpar}
R_0 \equiv \frac{21 f_1}{f_3}\,.
\ee

For the stability of the bosonic embedding given above we 
require the condition $F'(X)<0$, which translates into 
$f_3 X^2-2f_2 X+f_1>0$ \cite{Ketov11}. In order for this relation 
to hold for any real value of $X$, we have to require the 
condition\footnote{The stronger condition $f_2^2 \ll f_1 f_3$
was used in Ref.~\cite{Ketov11}.}
\be
f_2^2 <f_1 f_3\,.
\label{con1}
\ee
For the successful inflation the second term on the r.h.s. of 
Eq.~(\ref{Flag}) needs to be suppressed relative to the third term.
{}From the amplitude of the temperature anisotropies 
observed in CMB we require \cite{Ketov11} 
\be
f_3 \gg 1\,.
\label{con2}
\ee
In order to avoid large quantum corrections,
the scalaron mass in the low-curvature regime 
[defined in Eq.~(\ref{masslow})] needs to be smaller 
than the Planck mass, so that \cite{Ketov11}
\be
f_2 ^2 \gg f_1\,.
\label{con3}
\ee

There are two asymptotic regimes characterized by 
(A) $|R| \gg R_0$ and (B) $|R| \ll R_0$.
The former one is the high-curvature regime related 
to the generation of large-scale temperature anisotropies
observed in CMB.
The latter one is the low-curvature regime related to
reheating after inflation.
In the following we derive the approximate expression 
of $X$, as well as the Lagrangian $L$ 
in those two different regimes, improving the results of
Ref.~\cite{Ketov11}.

\begin{itemize}
\item (A) $|R| \gg R_0$

In this case the first and third terms on the l.h.s. of Eq.~(\ref{Xeq})
correspond to the dominant contributions, so that the 0-th order
solution satisfies $X_0^3+(R+R_0)X_0/30=0$.
In order to connect this solution to the one in the 
regime $|R| \ll R_0$ we require $X_0<0$, so that 
\be
X_0=-\sqrt{-\frac{R+R_0}{30}}\,.
\label{X0}
\ee
We regard the second and fourth terms in Eq.~(\ref{Xeq})
as the perturbations to the leading-order solution (\ref{X0}).
Setting $X=X_0+\delta X$ ($|\delta X| \ll |X_0|$) and expanding 
Eq.~(\ref{Xeq}) up to the linear order in $\delta X$, 
the perturbation $\delta X$ can be expressed in terms 
of $f_2$. This process leads to the following solution:
\ba
X &=& -\sqrt{-\frac{R+R_0}{30}} \nonumber \\
&+& \frac{f_2}{2} \frac{13R+33R_0}
{20f_3 (R+R_0)-33f_2 \sqrt{-30(R+R_0)}}.
\ea
Substituting this relation into Eq.~(\ref{bosof}) and
expanding it at linear order in $f_2$ yields
\ba
L &=& -\frac{1}{10} f_1 R+\frac{1}{180} 
f_3 (R^2+R_0^2)
\nonumber \\
& & -\frac{f_2}{900} (9R-11R_0) \sqrt{-30(R+R_0)}\,.
\label{Linf}
\ea
In the limit $|R| \gg R_0$ this Lagrangian 
reduces to 
\be
\hspace{0.5cm}
L \simeq -\frac{1}{10} f_1 R+\frac{1}{180} 
f_3 R^2+\frac{\sqrt{30}}{100} f_2 (-R)^{3/2}\,.
\label{laghigh}
\ee

\item (B) $|R| \ll R_0$

In this regime the dominant contributions in Eq.~(\ref{Xeq}) 
correspond to the third and fourth terms, so that 0-th 
order solution is given by 
\be
X_0=\frac{f_2 R}{f_3 (R+R_0)}\,,
\ee
which is negative.
Taking into account the first and second terms 
as the perturbations to $X_0$, we obtain 
the following solution: 
\ba
X &=& \frac{f_2 R}{f_3 (R+R_0)} \nonumber \\
&\times& \left[ 1+\frac32
\frac{f_2^2 R (13R+33R_0)}{f_3^2 (R+R_0)^3
-9f_2^2 R (R+11R_0)} \right].
\ea
Substituting this solution into Eq.~(\ref{bosof})
and expanding it in terms of $f_2$
up to the order of $f_2^4$ yields 
\be
\hspace{0.7cm}
L = -\frac13 f_1 R+\frac{f_2^2 R^2}{3f_3 (R+R_0)}
+\frac{f_2^4 (6R+11R_0) R^3}{f_3^3 (R+R_0)^4}.
\label{laglow}
\ee
\end{itemize}

Note that there is also the intermediate regime characterized 
by $|R+R_0| \ll R_0$, i.e. when $R$ is close to $-R_0$ \cite{Ketov11}.
In this case the first and fourth terms in Eq.~(\ref{Xeq}) 
are the dominant contributions.
Picking up the next-order contributions as well, we have 
\be
X=\left( \frac{f_2}{30f_3} \right)^{1/3} (-R_0)^{1/3}
+\frac{11f_2}{20f_3}\,.
\ee
In this regime the Lagrangian (\ref{bosof}) includes the term 
proportional to $R$ alone.
This transient era is followed by the reheating epoch 
characterized by the Lagrangian (\ref{laglow}).

In order to recover the standard behaviour of General Relativity  
in the low-energy regime we require that 
$f_1=3M_{\rm pl}^2/2$.
The mass squared of the scalar degree of 
freedom is given by $m^2=1/(3f''(R))$, 
where $f(R)$ is related 
with $L(R)$ as $L(R)=-M_{\rm pl}^2 f(R)/2$.
In the limit $|R| \ll R_0$ we have  
\be
m^2=\frac{21f_1 M_{\rm pl}^2}{4f_2^2}
=\frac{63M_{\rm pl}^4}{8f_2^2}\,.
\label{masslow}
\ee
In the high-curvature regime given by the Lagrangian 
(\ref{laghigh}) the scalaron mass squared is 
\be
M^2=\frac{15M_{\rm pl}^2}{f_3}\,.
\ee
Then the constants $f_{1,2,3}$ can be expressed by using 
the three mass scales $M_{\rm pl}$, $m$, and $M$, as
\be \label{repar}
f_1=\frac{3}{2} M_{\rm pl}^2\,,\quad
f_2=\sqrt{\frac{63}{8}} \frac{M_{\rm pl}^2}{m}\,,\quad
f_3=\frac{15M_{\rm pl}^2}{M^2}\,.
\ee
The conditions (\ref{con1}), (\ref{con2}), and (\ref{con3})
translate into 
\be
m>\sqrt{\frac{7}{20}}\,M\,,\qquad
M  \ll M_{\rm pl}\,,\qquad
m \ll M_{\rm pl}\,,
\label{mcon}
\ee
respectively.

\section{Inflationary dynamics}
\label{infsec}

Let us study the dynamics of inflation for the $f(R)$ model 
introduced in Sec.~\ref{modelsec}.
Here we are interested in the high-energy regime (A)
satisfying the condition $|R| \gg R_0$.

We consider the flat Friedmann-Lema\^{i}tre-Robertson-Walker
(FLRW) background described by the line element 
$ds^2=dt^2-a^2(t) d{\bm x}^2$, where $a(t)$ is the scale factor 
with the cosmic time $t$. The Ricci scalar is given by $R=-6(2H^2+\dot{H})$, 
where $H=\dot{a}/a$ is the Hubble parameter (the dots stand for the
derivatives with respect to $t$).

In order to study the dynamics of inflation, it is convenient to 
introduce the following dimensionless functions:
\be
\alpha \equiv \frac{M^2}{mH}\,,\qquad
\beta \equiv \frac{M^2}{H^2}\, .
\label{albe}
\ee
By using Eqs.~(\ref{rpar}) and (\ref{repar}), $R_0$ can be expressed 
as $R_0=21M^2/10$. During inflation the functions (\ref{albe}) should
satisfy the conditions $\alpha \ll 1$ and $\beta \ll 1$ (see below).
In Eq.~(\ref{Linf}) the term $f_3 R^2/180$ is the dominant 
contribution during inflation. Hence, we neglect the higher-order terms
beyond that of the first (linear) order in $\alpha$ and $\beta$.
Then the Lagrangian $f(R)=-2L(R)/M_{\rm pl}^2$ 
following from (\ref{Linf}) is given by\footnote{There are also some 
other approaches that give rise to the inflationary solution similar to
that in the Starobinsky's $f(R)$ model.
Those include a large non-minimal coupling to the inflaton 
field \cite{nonmini} and a macroscopic theory of the quantum vacuum 
in terms of conserved relativistic charges \cite{Klink}.} 
\be
f(R) \simeq \frac{3}{10}R-\frac{R^2}{6M^2}
-\frac{3\sqrt{105}}{100} \frac{(-R)^{3/2}}{m}\,.
\label{fRinf}
\ee
We assume that the Lagrangian (\ref{fRinf}) 
is valid by the end of inflation.

In the flat FLRW spacetime the field equations 
of motion are\footnote{Compared to the field equations 
given in Ref.~\cite{DTreview} we only need to change
$R \to -R$ and $f \to -f$, while ${\cal F}$ is unchanged.}
\ba
3 {\cal F} H^2 &=& (f-R {\cal F})/2-3H \dot{\cal F}\,,
\label{be1} \\
-2{\cal F} \dot{H} &=& \ddot{{\cal F}}-H \dot{\cal F}\,,
\label{be2}
\ea
where ${\cal F} \equiv f'(R)$.
We introduce the slow-roll parameters \cite{DTreview}
\be
\epsilon_1 \equiv -\frac{\dot{H}}{H^2}\,,\qquad
\epsilon_2 \equiv \frac{\dot{{\cal F}}}{2H{\cal F}}\,,\qquad
\epsilon_3 \equiv \frac{\ddot{{\cal F}}}{H \dot{{\cal F}}}\,,
\ee
which satisfy $|\epsilon_i| \ll 1$ ($i=1,2,3$).
{}From Eq.~(\ref{be2}) it follows that 
\be
\epsilon_1=-\epsilon_2 (1-\epsilon_3)\,.
\label{ep123}
\ee
In the following we carry out the linear expansion in terms of 
the variables $\epsilon_i$ ($i=1,2,3$), $\alpha$, $\beta$, and 
$s \equiv \ddot{H}/(H \dot{H})$.

For the Lagrangian (\ref{fRinf}) we have
\ba
{\cal F} &=& \frac{4H^2}{M^2} \left( 1+\frac{27 \sqrt{35}}{400} \alpha
+\frac{3}{40} \beta -\frac12 \epsilon_1 \right)\,,\label{Fva} \\
\dot{\cal F} &=& -\frac{8H^3}{M^2} \epsilon_1
\left( 1+\frac{27 \sqrt{35}}{800} \alpha
+\frac14 s \right)\,.
\ea
Then the variable $\epsilon_2$ is given by 
\be
\epsilon_2=-\epsilon_1 \left( 1-\frac{27 \sqrt{35}}{800} \alpha
-\frac{3}{40} \beta+\frac12 \epsilon_1 +\frac14 s \right)\,.
\ee
Comparing this with Eq.~(\ref{ep123}), we obtain
\be
\epsilon_3=-\frac{27 \sqrt{35}}{800} \alpha
-\frac{3}{40} \beta+\frac12 \epsilon_1 +\frac14 s\,.
\label{ep3d}
\ee
Similarly, Eq.~(\ref{be1}) gives the following relations:
\be
\epsilon_1=\frac{3\sqrt{35}}{200} \alpha+\frac{1}{20} \beta\,,
\label{ep1}
\ee
and 
\be
\epsilon_2=-\frac{3\sqrt{35}}{200} \alpha-\frac{1}{20} \beta\,.
\label{ep2}
\ee

Equation (\ref{ep1}) is equivalent to 
\be
\dot{H}=-\frac{3\sqrt{35}}{200} \frac{M^2}{m}
\left( H+\frac{10m}{3\sqrt{35}} \right)\,.
\ee
This differential equation can be easily integrated. 
It yields 
\be
H(t)=\left( H_i +\frac{10m}{3\sqrt{35}} \right)
\exp \left[ \frac{3\sqrt{35}}{200} \frac{M^2}{m}
(t_i-t) \right]-\frac{10m}{3\sqrt{35}},
\label{Hana}
\ee
where $H_i$ is the initial value of $H$ at $t=t_i$.
So we find
\be
s=-\frac{3\sqrt{35}}{200}\alpha\,.
\label{s}
\ee
Substituting Eqs.~(\ref{ep1}) and (\ref{s}) into 
Eq.~(\ref{ep3d}), we obtain
\be
\epsilon_3=-\frac{3\sqrt{35}}{100} \alpha
-\frac{1}{20} \beta\,.
\label{ep3}
\ee

The end of inflation ($t=t_f$) is identified
by the condition $\epsilon_1=1$.
By using the solution (\ref{Hana}), we have
\ba
\hspace{-0.6cm}
t_i-t_f &=& \frac{200m}{3\sqrt{35}M^2}
\ln \biggl(
\frac{63M^2}{80m (3\sqrt{35}H_i+10m)}
\nonumber \\
\hspace{-0.6cm}  &\times&
\biggl[ 1+\frac{800}{63} \left( \frac{m}{M} \right)^2
+\sqrt{1+\frac{1600}{63} \left( \frac{m}{M} \right)^2}
\biggr] \biggr).
\label{tfi}
\ea
We define the number of e-foldings from the onset of inflation ($t=t_i$)
to the end of inflation ($t=t_f$) as $N (t_i) \equiv \int_{t_i}^{t_{f}}H\,dt$.
{}From Eqs.~(\ref{Hana}) and (\ref{tfi}) we can express $N(t_i)$ in 
terms of $H_i$, $M$, and $m$.
The number of e-foldings $N$ corresponding to the time $t$ 
can be derived by replacing $H_i$ in the expression of $N(t_i)$ 
for $H$. It follows that 
\ba
N &=& \frac{1}{126\alpha^2}
\biggl[ 3\alpha (80\sqrt{35}-21\alpha-\sqrt{7(63\alpha^2+1600\beta)})
\nonumber \\
&-&400 \beta (8 \ln 2+3 \ln 5)+800\beta 
\nonumber \\
&\times& \ln \biggl( \frac{\sqrt{7} (63\alpha^2+800\beta)+
21\alpha\sqrt{63\alpha^2+1600\beta}}{21\alpha+2\sqrt{35} \beta}
\biggr) \biggr].\nonumber \\
\label{efold}
\ea

In the limit $\alpha \to 0$ one has $N \to 10/\beta-1/2$, i.e.
$\beta \to 20/(2N+1)$.
In this case the $R^2/(6M^2)$ term in the Lagrangian (\ref{fRinf})
dominates over the dynamics of inflation, which corresponds to the
Starobinsky's $f(R)$ model \cite{Starobinsky80}. 
In another limit $\beta \to 0$ 
it follows that  $N \to 40\sqrt{35}/(21\alpha)-1$, i.e.
$\alpha \to 40\sqrt{35}/[21(N+1)]$.
Then we obtain the following bounds on 
$\alpha$ and $\beta$:
\be
0<\alpha<\frac{40\sqrt{35}}{21(N+1)}\,,
\qquad
0<\beta<\frac{20}{2N+1}\,.
\label{alcon}
\ee

In order to realize inflation with $N=60$, 
for example, the two variables need to be in the range
$0<\alpha<0.185$ and $0<\beta<0.165$.
For the number of e-foldings relevant to the CMB temperature anisotropies 
($50 \lesssim N \lesssim 60$) the slow-roll parameters
given in Eqs.~(\ref{ep1}), (\ref{ep2}), (\ref{ep3}) are much 
smaller than unity, so that the slow-roll approximation
employed above is justified.

\section{Observational tests}
\label{obsersec}

Let us study whether the $f(R)$ model (\ref{fRinf}) can satisfy 
observational constraints of the CMB temperature anisotropies.
The power spectra of scalar and tensor perturbations generated
during inflation based on $f(R)$ theories have been calculated 
in Ref.~\cite{spectra} (see also Ref.~\cite{DTreview}).

The scalar power spectrum of the curvature perturbation 
is given by \cite{DTreview}
\be
{\cal P}_{\rm s}=\frac{1}{24\pi^2 {\cal F}}
\left( \frac{H}{M_{\rm pl}} \right)^2
\frac{1}{\epsilon_2^2}\,.
\label{Ps}
\ee
Using Eqs.~(\ref{Fva}), (\ref{ep1}), and (\ref{ep2}), 
it follows that
\be
{\cal P}_{\rm s} \simeq \frac{1250}{3\pi^2}
\left( \frac{M}{M_{\rm pl}} \right)^2
\left( 3\sqrt{35} \alpha+10\beta \right)^{-2}\,,
\ee
where, in the expression of ${\cal F}$, we have neglected 
the terms $\alpha$ and $\beta$ relative to 1.
Using the WMAP normalization ${\cal P}_{\rm s}=2.4 \times 10^{-9}$
at the pivot wave number $k_0=0.002$~Mpc$^{-1}$ \cite{WMAP7}, 
the mass $M$ is constrained to be 
\be
M \simeq 7.5 \times 10^{-6} \left( 3\sqrt{35} \alpha+10\beta
\right) M_{\rm pl}\,.
\label{WMAPnor}
\ee
In the limit that $\alpha \to 0$ and $\beta \to 20/(2N+1)$
we have $M/M_{\rm pl}=7.5 \times 10^{-4}/(N+1/2)$.
In another limit $\alpha \to 40\sqrt{35}/[21(N+1)]$ 
and $\beta \to 0$ it follows that 
$M/M_{\rm pl}=1.5 \times 10^{-3}/(N+1)$.
In the intermediate regime characterized by (\ref{alcon})
we can numerically find the values of $\alpha$ and $\beta$ 
for given $N$ satisfying the constraint (\ref{efold}), 
which allows us to evaluate $M$ from Eq.~(\ref{WMAPnor}).
{}From Eq.~(\ref{albe}) the mass scale $m$ is also
known by the relation $m=(\sqrt{\beta}/\alpha)M$.

\begin{figure}
\includegraphics[height=3.2in,width=3.3in]{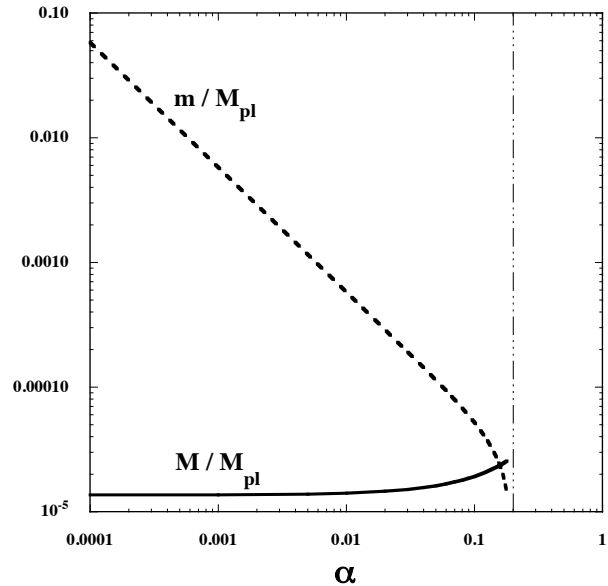}
\caption{The two masses $M$ and $m$ 
versus the variable $\alpha$ in the regime 
$10^{-4} \le \alpha \le 0.18$ 
for the number of e-foldings $N=55$. 
We also show the upper bound $\alpha_{\rm max}=0.201$
determined by Eq.~(\ref{alcon}).
$M$ is weakly dependent on $\alpha$
with the order of $10^{-5}M_{\rm pl}$, whereas
$m$ strongly depends on $\alpha$.
The condition $m>\sqrt{7/20}\,M$ is satisfied
for $\alpha<0.178$.
\label{fig1}}
\end{figure}

In Fig.~\ref{fig1} we plot $M$ and $m$ versus $\alpha$
in the regime $10^{-4} \le \alpha \le 0.18$ for $N=55$. 
In this case $\alpha$ is bounded to be 
$0<\alpha<0.201$ from Eq.~(\ref{alcon}).
The mass $M$ weakly depends on $\alpha$ with 
the order of $10^{-5}M_{\rm pl}$, while $m$ changes 
significantly depending on the values of $\alpha$.
For $\alpha$ much smaller than 1 we have $m \gg M$, 
while $m$ is of the same order as $M$ for $\alpha \gtrsim 0.1$.
We recall that there is the condition $m>\sqrt{7/20}\,M$
coming from the requirement (\ref{con1}).
For $N=55$ this condition gives the upper 
bound $\alpha<0.178$.

The scalar spectral index $n_{\rm s}$ is defined by 
$n_{\rm s}=1+d \ln {\cal P}_{\rm s}/d \ln k$, which
is evaluated at the Hubble radius crossing $k=aH$ 
(where $k$ is a comoving wave number) \cite{Lidsey, Bassett}.
In $f(R)$ gravity this is given by \cite{DTreview}
\be
n_{\rm s}=1-4\epsilon_1+2\epsilon_2-2\epsilon_3\,.
\ee
On using Eqs.~(\ref{ep1}), (\ref{ep2}) and (\ref{ep3}), 
we obtain
\be
n_{\rm s} =
1-\frac{3\sqrt{35}}{100}\alpha-\frac15 \beta\,.
\label{nsab}
\ee

The tensor power spectrum is given by \cite{DTreview}
\be
{\cal P}_{\rm t}=\frac{2}{\pi^2 {\cal F}}
\left( \frac{H}{M_{\rm pl}} \right)^2\,.
\label{Pt}
\ee
{}From Eqs.~(\ref{Ps}) and (\ref{Pt}) the tensor-to-scalar
ratio is 
\be
r \equiv \frac{{\cal P}_{\rm t}}{{\cal P}_{\rm s}}
=48\epsilon_2^2=\frac{3}{2500}
\left( 3\sqrt{35} \alpha+10 \beta \right)^2\,.
\label{rab}
\ee

In the limit $\alpha \to 0$ and $\beta \to 20/(2N+1)$
the observables (\ref{nsab}) and (\ref{rab}) reduce to 
\ba
n_{\rm s} (\alpha \to 0) &=& 1-\frac{4}{2N+1}\,,
\label{asy1n} \\
r (\alpha \to 0) &=& \frac{48}{(2N+1)^2}\,,
\label{asy1}
\ea
which agree with those in the Starobinsky's
$f(R)$ model \cite{spectra}. 
For $N=55$ one has $n_{\rm s} (\alpha \to 0)=0.964$
and $r (\alpha \to 0)=3.896 \times 10^{-3}$.
In another limit $\alpha \to 40\sqrt{35}/[21(N+1)]$ 
and $\beta \to 0$ it follows that 
\ba
n_{\rm s} (\beta \to 0) &=& 1-\frac{2}{N+1}\,,
\label{asy2n} \\
r (\beta \to 0) &=& \frac{48}{(N+1)^2}\,.
\label{asy2}
\ea
For $N=55$ one has $n_{\rm s} (\beta \to 0)=0.964$
and $r (\beta \to 0)=1.531 \times 10^{-2}$.
While the scalar spectral indices (\ref{asy1n}) and 
(\ref{asy2n}) are practically identical for $N \gg 1$, 
$r (\beta \to 0)$ is about four times as large as $r(\alpha \to 0)$.
For the intermediate values of $\alpha$ between 0 and 
$40\sqrt{35}/[21(N+1)]$ we need to numerically derive $\beta$ 
satisfying Eq.~(\ref{efold}) for given $N$.
Then $n_{\rm s}$ and $r$ are known from 
Eqs.~(\ref{nsab}) and (\ref{rab}).

\begin{figure}
\includegraphics[height=3.3in,width=3.5in]{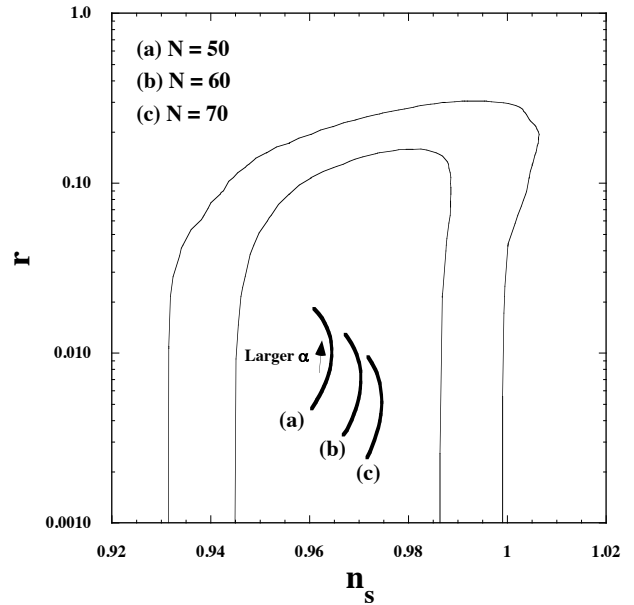}
\caption{The three thick lines show the theoretical values 
of $n_{\rm s}$ and $r$ for $N=50, 60, 70$ with
$\alpha$ ranging in the region (\ref{alcon}).
The thin solid curves are the 1$\sigma$ (inside)
and 2$\sigma$ (outside) observational contours 
constrained by the joint data analysis of 
WMAP7, BAO, and HST.
For $\alpha \to 0$,  $n_{\rm s}$ and $r$ are given by 
Eqs.~(\ref{asy1n}) and (\ref{asy1}).
In the limit $\beta \to 0$,
$n_{\rm s}$ and $r$ approach the values given in 
Eqs.~(\ref{asy2n}) and (\ref{asy2}).
\label{fig2}}
\end{figure}

In Fig.~\ref{fig2} we plot the theoretical values of $n_{\rm s}$
and $r$ in the $(n_{\rm s}, r)$ plane for $N=50, 60, 70$ together 
with the $1\sigma$ and $2\sigma$ observational contours
constrained by the joint data analysis of WMAP7 \cite{WMAP7}, 
Baryon Acoustic Oscillations (BAO) \cite{BAO}, 
and the Hubble constant measurement (HST) \cite{HST}.
The observational bounds are derived by using the 
standard consistency relation $r=-8n_{\rm t}$ \cite{Bassett}, 
where $n_{\rm t}=d \ln {\cal P}_{\rm t}/d \ln k$ is the tensor 
spectral index. In $f(R)$ gravity this relation also holds
by using the equivalence of the power spectra between 
the Jordan and Einstein frames \cite{DTreview}
(see also Refs.~\cite{equiva}).

The Starobinsky's $f(R)$ model, which corresponds to the 
limit $\alpha \to 0$ with the observables 
given in Eqs.~(\ref{asy1n}) and (\ref{asy1}), 
is well within the current observational 
bound. In the regime $\alpha \ll \beta$ one has 
$\beta \simeq 20/(2N+1)-\sqrt{35}\alpha/5$, so that 
\ba
\hspace{-0.3cm}
n_{\rm s} (\alpha \ll \beta) &=& 1-\frac{4}{2N+1}
+\frac{\sqrt{35}}{100} \alpha\,,\\
\hspace{-0.3cm}
r (\alpha \ll \beta) &=& \frac{48}{(2N+1)^2}
\left[ 1+\frac{\sqrt{35}(2N+1)}{200} \alpha \right]^2\,.
\ea
This shows that both $n_{\rm s}$ and $r$ increase for
larger $\alpha$ satisfying the condition $\alpha \ll \beta$. 
As we see in Fig.~\ref{fig2}, $n_{\rm s}$ switches to 
decrease at some value of $\alpha$, whereas
$r$ continuously grows toward the asymptotic value 
given in Eq.~(\ref{asy2}).

{}From Fig.~\ref{fig2} we find that the $f(R)$ model (\ref{fRinf})
in which $\alpha$ is in the range (\ref{alcon}) is inside the 
$1\sigma$ observational contour.
The condition $m>\sqrt{7/20}\,M$ provides the 
constraints $\alpha<0.194$, $\alpha<0.165$, $\alpha<0.143$
for $N=50, 60, 70$ respectively, while the bound (\ref{alcon})
in each case corresponds to 
$\alpha<0.221$, $\alpha<0.185$, $\alpha<0.159$.
When $N=60$ the scalar spectral index and the tensor-to-scalar ratio 
are $n_{\rm s}=0.969$, $r=0.0110$ for $\alpha=0.165$ and 
$n_{\rm s}=0.967$, $r=0.0129$ for $\alpha=0.185$, which
are not very different from each other.
For the background in which inflation is sustained with the 
number of e-foldings $N>50$ the model is consistent 
with the current observations.

Note that the nonlinear parameter $f_{\rm NL}$ of the
scalar non-Gaussianities is of the order of the slow-roll 
parameters in $f(R)$ gravity \cite{Defe}.
Hence, in current observations, this does not provide 
additional constraints to those studied above.

\section{Effective scalar potential}
\label{scalarpotensec}

In this section we derive the effective scalar potential and the kinetic terms 
of a {\it complex} scalaron field in the low-energy regime (B) 
characterized by $|R| \ll R_0$. 
In doing so, let us return to the original $F(\car)$  supergravity action 
(\ref{frsup}) and perform the superfield Legendre transformation \cite{kkw}. 
We temporarily set $M_{\rm pl}= 1$ to simplify our calculations. 
It yields the equivalent action 
\be 
S=\int d^4 x\,d^2 \theta\, {\cal E}\, \left[ -\cy \car +Z(\cy) \right] 
+{\rm H.c.}\,,
\label{frsup2}
\ee
where we have introduced the new covariantly chiral superfield $\cy$ and the 
new holomorphic function $Z(\cy)$ related to the function $F$ as 
\be \label{zfun}
F(\car) = -\car \cy(\car) + Z(\cy(\car))\,.
\ee
The equation of motion of the superfield $\cy$, which follows from the variation 
of the action (\ref{frsup2}) with respect to $\cy$, has the algebraic form
\be \label{yem}
\car = Z'(\cy) \, ,
\ee
so that the function $\cy(\car)$ is obtained by inverting the function $Z'$.  
Substituting the solution $\cy(\car)$ back into the action (\ref{frsup2}) yields 
the original action (\ref{frsup}) because of Eq.~(\ref{zfun}). 
We also find 
\be \label{leg}
\cy =- F'(\car) \, .
\ee
The inverse function $\car(\cy)$ always exists under the physical condition  
$F'(\car) \neq 0$. As regards the $F$-function (\ref{Flag}), Eq.~(\ref{leg}) yields
a quadratic equation with respect to $\car$, whose solution is
\be \label{quads}
\car(\cy) =  \frac{\sqrt{14}M^2}{20m} \left[ 
1 - \sqrt{ 1 + \frac{80m^2}{21M^2}({\cal Y}-3/4) }
\right]\,,
\ee 
where we have used the parametrization (\ref{repar}). 
Equation (\ref{quads}) is also 
valid for the leading complex scalar field 
components $\left.\car\right|=\bar{B}/3=\bar{X}$ and 
$\left.\cy\right|\equiv Y$, where $Y$ is the complex scalaron field.

The kinetic terms of $\cy$ are obtained by using the identity 
\be \label{siegel}  
\int d^4 x\,d^2 \theta\, {\cal E}\, \cy \car  +{\rm H.c.}=
\int d^4 x \, d^4 \theta \, E^{-1} (\cy +\bar{\cy} ) \, ,
\ee
where $E^{-1}$ is the full curved superspace density \cite{SGbook}.
 Therefore, the K\"ahler potential reads 
\be \label{kahl}
K = -3 \ln \left( \cy + \bar{\cy} \right) \, .
\ee
It gives rise to the kinetic terms
\ba \label{kinc}
{\cal L}_{\rm kin} &=& \left. \frac{\pa^2 K}{\pa\cy \pa\bar{\cy}}\right|_{\cy=Y}
\pa_{\mu}Y\pa^{\mu}\bar{Y} \nonumber \\
&=& 3 \frac{ \pa_{\mu}Y\pa^{\mu}\bar{Y} }
{(Y +\bar{Y})^2}=
3 \frac{ (\pa_{\mu}y)^2 +  (\pa_{\mu}z)^2}{4y^2} \; ,
\ea
where we have used the notation $Y=y +iz$ in terms of the two real fields 
$y$ and $z$. The imaginary component $z$ corresponds to 
a pseudo-scalar field.
The kinetic terms (\ref{kinc}) represent 
the {\it non-linear sigma model} \cite{ketovbook}  with 
the hyperbolic target space of (real) dimension two, whose metric is known as the 
standard 
Poincar\'e metric. The kinetic terms are invariant under arbitrary rescalings 
$Y \to AY$ 
with constant parameter $A \neq 0$. 

The effective scalar potential $V(Y,\bar{Y})$ of a complex scalaron $Y$ 
in the regime (B), where supergravity decouples 
(it corresponds to rigid supersymmetry), is easily derived from
Eq.~(\ref{frsup2}) when keeping only scalars (i.e. ignoring their spacetime derivatives
together with all fermionic contributions) and 
eliminating the auxiliary fields, near the minimum of the scalar potential.
We find 
\be \label{scals}
V = \frac{21}{2} \left|  Z'(Y) \right|{}^2 =  \frac{21}{2} \left|  
\car(Y) \right|{}^2 \, ,
\ee
which gives rise to the chiral superpotential
\be \label {chipot} 
W (\cy) = \sqrt{\frac{21}{2}} Z(\cy)\,.
\ee
The superfield equations (\ref{kahl}) and (\ref{chipot}) are model-independent, i.e. 
they apply to any function $F({\cal R})$ in the large $M_{\rm pl}$ limit, 
near the minimum of the scalar potential with the vanishing cosmological constant.  

There is no field redefinition that would bring all the kinetic terms (\ref{kinc}) to 
the 
free form. The canonical (free) kinetic term of a {\it real} scalaron $y$ alone can be 
obtained via the field redefinition 
\be \label{repara}
y = A \exp( -\sqrt{2/3}\,\phi)\, .
\ee
The scalaron potential vanishes at $y=3/4$.
Demanding that this minimum corresponds to $\phi=0$, we have  
$A=3/4$ and hence $y=(3/4) \exp( -\sqrt{2/3}\,\phi)$.
Defining a rescaled field $\chi$ as $\chi=\sqrt{8/3}\,z$, the 
kinetic term (\ref{kinc}) can be written as 
\be \label{kine2}
{\cal L}_{\rm kin}=\frac12 (\partial_{\mu} \phi)^2+
\frac12 e^{2\sqrt{2/3}\,\phi/M_{\rm pl}} (\partial_{\mu} \chi)^2\,.
\ee
Here and in what follows we restore the reduced Planck
mass $M_{\rm pl}$.

The total potential (\ref{scals}) including both the fields
$\phi$ and $\chi$ is given by 
\be \label{Vpc}
V(\phi, \chi)=\frac{147M^4M_{\rm pl}^2}{400m^2} \left|
\sqrt{{\cal B}(\phi)+i{\cal C}(\chi)}-1 \right|^2\,, 
\ee
where 
\ba  \label{BC}
{\cal B}(\phi) &=& 1+\frac{20m^2}{7M^2}
\left( e^{-\sqrt{2/3}\,\phi/M_{\rm pl}}-1  \right)\,,\\
{\cal C}(\chi) &=& \frac{80m^2}{21M^2}
\sqrt{\frac38}\,\frac{\chi}{M_{\rm pl}}\,.
\ea
In order to express (\ref{Vpc}) in a more convenient form
we write $\sqrt{{\cal B}(\phi)+i{\cal C}(\chi)}=p+iq$, where 
$p$ and $q$ are real.
This gives the relations $p^2-q^2={\cal B}(\phi)$ and 
$2pq={\cal C}(\chi)$. Solving these equations for $p$, 
we find
\be 
p=\frac{1}{\sqrt{2}} \left[ {\cal B}(\phi)+
\sqrt{{\cal B}^2(\phi)+{\cal C}^2(\chi)}
\right]^{1/2}\,,
\ee
where we have chosen the solution $p>0$ to recover
$p=\sqrt{{\cal B}(\phi)}$ for ${\cal B}(\phi)>0$
in the limit ${\cal C}(\chi) \to 0$.
Then the field potential (\ref{Vpc}) reads 
\ba 
\hspace{-0.3cm}V(\phi, \chi) &=& 
\frac{147M^4 M_{\rm pl}^2}{400m^2} \biggl[1+
\sqrt{{\cal B}^2 (\phi)+{\cal C}^2 (\chi)} \nonumber \\
\hspace{-0.3cm}& &-\sqrt{2} \left\{ {\cal B}(\phi)
+\sqrt{{\cal B}^2 (\phi)+{\cal C}^2 (\chi)} \right\}^{1/2} \biggr].
\label{Vtotal}
\ea

In the absence of the pseudo-scalar $\chi$
the potential (\ref{Vtotal}) reduces to 
\be  \label{scalpot}
V(\phi)= \frac{147M^4 M_{\rm pl}^2}{400m^2}
\left[ 1+|{\cal B}(\phi)|-\sqrt{2} \left\{ {\cal B}(\phi)
+|{\cal B} (\phi)| 
\right\}^{1/2} \right].
\ee
For the field $\phi$ satisfying the condition 
${\cal B}(\phi)<0$ it follows that 
\be  \label{scalpot1}
V(\phi)=\frac{21}{20}M^2 M_{\rm pl}^2
\left( 1-e^{-\sqrt{2/3}\,\phi/M_{\rm pl}}
\right)\,,
\ee
which approaches the constant 
$V(\phi) \to 21M^2M_{\rm pl}^2/20$
in the limit $\phi \to \infty$.
Defining the slow-roll parameter 
$\epsilon_V=(M_{\rm pl}^2/2)(V_{,\phi}/V)^2$, 
we have
\be  
\epsilon_V=\frac{x^2}{3(1-x)^2}\,,\qquad
x=e^{-\sqrt{2/3}\,\phi/M_{\rm pl}}\,.
\ee
The end of inflation is characterized by the criterion $\epsilon_V=1$. 
This gives $x_f=e^{-\sqrt{2/3}\,\phi_f/M_{\rm pl}}=(3-\sqrt{3})/2$ 
and hence $\phi_f=0.558M_{\rm pl}$.
For $m>M$ the condition ${\cal B}(\phi_f)<0$
is satisfied, so that the potential (\ref{scalpot1}) is valid 
at the end of inflation.
If $m$ is close to the border value $\sqrt{7/20}\,M$ 
---see Eq.~(\ref{mcon})---, then the potential (\ref{scalpot1}) 
is already invalid at the end of inflation.

For small $\phi$ satisfying the condition 
${\cal B}(\phi)>0$ the potential (\ref{scalpot}) reads 
\be  \label{scalpot4}
\hspace{-0.2cm}V(\phi)=\frac{147M^4M_{\rm pl}^2}{400m^2} \left[ 1-\sqrt{1+
\frac{20m^2}{7M^2} \left( e^{-\sqrt{2/3}\phi/M_{\rm pl}}-1
\right)} \right]^2.
\ee
In this case Taylor expansion around $\phi=0$ gives 
rise to the leading-order contribution $V(\phi)=m^2 \phi^2/2$.
Reheating occurs around the potential minimum
through the oscillation of the canonical field $\phi$.

The total effective potential involving the interaction between 
the fields $\phi$ and $\chi$ is given by Eq.~(\ref{Vtotal}).
Expanding the potential (\ref{Vtotal}) around $\phi=\chi=0$
and picking up the terms up to fourth-order in the fields, 
we obtain 
\begin{widetext} 
\ba
V(\phi, \chi) 
&\simeq&  
\frac{1}{2}m^2 \phi^2+
\frac{\sqrt{6}m^2 (10m^2-7M^2)}{42M^2M_{\rm pl}}\phi^3
+\frac{(1500m^4-1260m^2M^2+343M^4)m^2}
{1764M^4 M_{\rm pl}^2}\phi^4 
+\frac{1}{2}m^2 \chi^2-\frac{25 m^6}{49M^4 M_{\rm pl}^2}\chi^4 
\nonumber \\
& & +\frac{5\sqrt{6}m^4}{21M^2M_{\rm pl}} \phi \chi^2
+\frac{5m^4 (10m^2-7M^2)}{147M^4M_{\rm pl}^2}\phi^2 \chi^2\,.
\label{Vre}
\ea
\end{widetext} 
The scalaron $\phi$ is coupled to to the pseudo-scalar $\chi$
through the interaction given in the second line of Eq.~(\ref{Vre}).

\section{Preheating after inflation}
\label{reheatingsec}

We study the dynamics of preheating for the two-field system 
described by the kinetic term (\ref{kine2}) and 
the effective potential (\ref{Vtotal}). The background equations 
of motion on the flat FLRW background are 
\ba 
& & 3M_{\rm pl}^2 H^2=\dot{\phi}^2/2+e^{2b}\dot{\chi}^2/2+V\,,
\label{Hubeq}\\
& & \ddot{\phi}+3H \dot{\phi}+V_{,\phi}-b_{,\phi}e^{2b} \dot{\chi^2}=0\,,
\label{phieqm} \\
& & \ddot{\chi}+(3H+2b_{,\phi} \dot{\phi})\dot{\chi}+e^{-2b}V_{,\chi}=0\,,
\label{ddotchi}
\ea
where $b(\phi)=\sqrt{2/3}\,\phi/M_{\rm pl}$ and 
``${}_{,\phi}$'' represents a partial derivative with 
respect to $\phi$.

In Fourier space the field perturbations $\delta \phi_k$ 
and $\delta \chi_k$ with the comoving wave number 
$k$ obey the following equations (see Refs.~\cite{multi}
for related works):
\ba 
& &\ddot{\delta \phi}_k+3H \dot{\delta \phi}_k+[k^2/a^2+V_{,\phi \phi}
-(2b_{,\phi}^2+b_{,\phi \phi}) e^{2b} \dot{\chi}^2]\delta \phi_k \nonumber \\
& &=-V_{,\phi \chi} \delta \chi_k+2b_{,\phi}e^{2b} \dot{\chi}
\dot{\delta \chi}_k \,,\label{delphieq} \\
& &\ddot{\delta \chi}_k+(3H+2b_{,\phi} \dot{\phi}) 
\dot{\delta \chi}_k
+(k^2/a^2+e^{-2b} V_{,\chi \chi})\delta \chi_k \nonumber \\
& &=
-e^{-2b} (V_{,\phi \chi}-2b_{,\phi}V_{,\chi}+2b_{,\phi \phi}
e^{2b} \dot{\phi} \dot{\chi})\delta \phi_k
-2b_{,\phi} \dot{\chi} \dot{\delta \phi}_k. \nonumber \\
\label{delchieq}
\ea
%

\begin{figure}
\includegraphics[height=2.6in,width=3.2in]{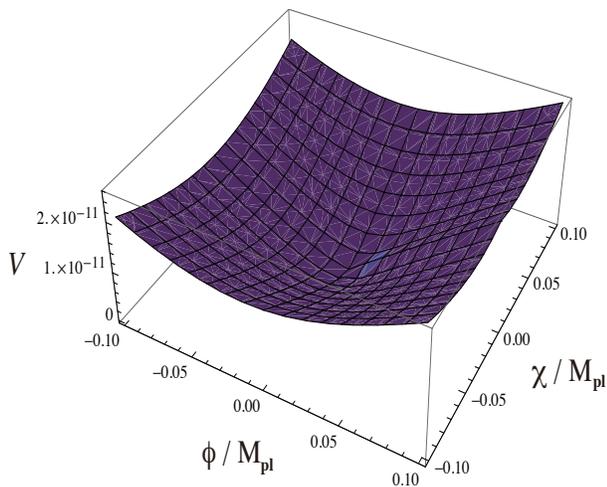}
\caption{
Effective potential (\ref{Vtotal}) for $m=1.14 \times 10^{-4}M_{\rm pl}$
and $M=1.62 \times 10^{-5} M_{\rm pl}$ in the region
$-0.1<\phi/M_{\rm pl}<0.1$ and $-0.1<\chi/M_{\rm pl}<0.1$.
\label{fig3}}
\end{figure}

\begin{figure}
\includegraphics[height=3.0in,width=3.1in]{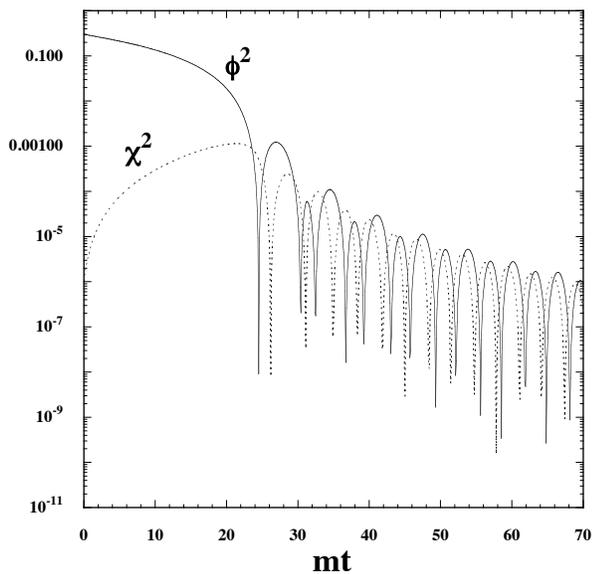}
\caption{
Evolution of the background fields $\phi^2$ and $\chi^2$
(both are normalized by $M_{\rm pl}^2$) for 
$m=1.14 \times 10^{-4} M_{\rm pl}$ and 
$M=1.62 \times 10^{-5} M_{\rm pl}$
with the initial conditions $\phi=0.55 M_{\rm pl}$, 
$\chi=10^{-3} M_{\rm pl}$, 
$\dot{\phi}=-1.6 \times 10^{-2}mM_{\rm pl}$, and 
$\dot{\chi}=1.5 \times 10^{-3}mM_{\rm pl}$.
\label{fig4}}
\end{figure}

The derivative $V_{,\chi}$ of the potential (\ref{Vtotal})
vanishes at $\chi=\pm \chi_c$, where
\be \label{chic}
\chi_c=\frac{\sqrt{210}M}{20m}
\left[ 1-e^{-\sqrt{2/3}\,\phi/M_{\rm pl}}-\frac{21}{80}
\left( \frac{M}{m} \right)^2 \right]^{1/2}M_{\rm pl}\,.
\ee
The local minima exist in the $\chi$ direction 
provided that 
\be \label{phic}
\phi>\sqrt{\frac32} \ln \left[ 1-\frac{21}{80}
\left( \frac{M}{m} \right)^2 \right]^{-1} M_{\rm pl}
\equiv \phi_c\,,
\ee
whereas they disappear for $\phi<\phi_c$.
In Fig.~\ref{fig3} we plot the potential (\ref{Vtotal})
with respect to $\phi$ and $\chi$ for 
$m=1.14 \times 10^{-4}M_{\rm pl}$
and $M=1.62 \times 10^{-5} M_{\rm pl}$.
Since $\phi_c=6.5 \times 10^{-3}M_{\rm pl}$ in this case, 
the potential has the local minima in the $\chi$ direction 
for $\phi>6.5 \times 10^{-3}M_{\rm pl}$.
{}From Eq.~(\ref{chic}) the field value $\chi_c$ increases
for larger $\phi$. For the model parameters used 
in Fig.~\ref{fig3}, for example, one has
$\chi_c=0.028 M_{\rm pl}$ at $\phi=0.1 M_{\rm pl}$ and  
$\chi_c=0.059 M_{\rm pl}$ at $\phi=0.5M_{\rm pl}$.

If the initial conditions of the fields are $0<\chi<\chi_c$ 
and $\phi>\phi_c$, the field $\chi$ grows 
toward the local minimum at $\chi=\chi_c$.
After $\phi$ drops below $\phi_c$, the field $\chi$
approaches the global minimum at $\chi=0$.
In Fig.~\ref{fig4} we show one example for the
evolution of the background fields $\phi$ and $\chi$
with the same values of $m$ and $M$ as those 
in Fig.~\ref{fig3}.
The energy density of the field $\chi$ catches up to 
that of the inflaton around the onset of reheating.

As we see in Eq.~(\ref{phic}), the critical field value $\phi_c$
gets smaller for increasing $m/M$. 
Hence, for larger $m/M$, the potential (\ref{Vtotal}) 
possesses the local minima at $\chi=\pm \chi_c$ 
for a wider range of $\phi$.
The potential in the region $|\chi|<\chi_c$ 
can be flat enough to lead to inflation by the slow-roll 
evolution of the field $\chi$, even if $\phi$ is smaller 
than $\phi_f=0.558M_{\rm pl}$.
For larger ratio $m/M$ inflation ends with the field value 
much smaller than $\phi_f$.
If $m/M=20$ and $m/M=83$, for example, the amplitudes of 
the field $\phi$ at the onset of oscillations are 
$\phi_i =1.5 \times 10^{-2} M_{\rm pl}$ and 
$\phi_i = 5.0 \times 10^{-3} M_{\rm pl}$, respectively.

Let us consider the regime where the condition 
\be \label{phicon}
\left( \frac{m}{M} \right)^2 \frac{|\phi|}{M_{\rm pl}} \ll 1
\ee
is satisfied. 
Then the potential (\ref{Vre}) is approximately
given by $V(\phi, \chi) \simeq m^2 \phi^2/2+m^2 \chi^2/2$, 
in which case both $\phi$ and $\chi$ have the 
same mass $m$.
This gives rise to the matter-dominated epoch 
(where $H=2/(3t)$) driven by the oscillations 
of two massive scalar fields.
{}From Eq.~(\ref{phieqm}) we have that  
$\ddot{\phi}+(2/t)\dot{\phi}+m^2 \phi \simeq 0$, 
whose solution is
\be
\phi(t) \simeq \frac{\pi}{2mt} \phi_i \sin (mt)\,.
\label{infosc}
\ee
Here the initial field value $\phi_i$ corresponds to the
time $t_i=\pi/(2m)$.

In order to discuss the dynamics of the field perturbations 
in Eqs.~(\ref{delphieq}) and (\ref{delchieq}) 
we define the two frequencies $\omega_{\phi}$ 
and $\omega_{\chi}$, as 
$\omega_{\phi}^2=k^2/a^2+V_{,\phi \phi}
-(2b_{,\phi}^2+b_{,\phi \phi}) e^{2b} \dot{\chi}^2$ and 
$\omega_{\chi}^2=k^2/a^2+e^{-2b}V_{,\chi \chi}$.
As long as the condition (\ref{phicon}) is satisfied, 
it is sufficient to pick up the terms up to cubic order in fields.
It then follows that 
\ba \label{omephi}
\omega_{\phi}^2 & \simeq &
\frac{k^2}{a^2}+m^2+\frac{\sqrt{6}m^2 (10m^2-7M^2)}
{7M^2 M_{\rm pl}} \phi\,,\\
\omega_{\chi}^2 &=&
\frac{k^2}{a^2}+m^2e^{-2b}+\frac{10\sqrt{6}m^4}
{21M^2 M_{\rm pl}}e^{-2b} \phi\,,
\ea
where, in Eq.~(\ref{omephi}), we have neglected the contribution 
of the term $-(2b_{,\phi}^2+b_{,\phi \phi}) e^{2b} \dot{\chi}^2$.

\begin{figure}
\includegraphics[height=3.0in,width=3.1in]{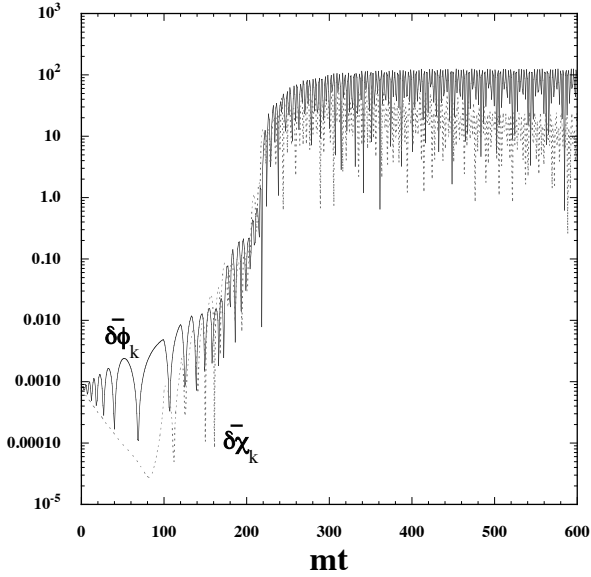}
\caption{
Evolution of the field perturbations
$\bar{\delta \phi}_k=k^{3/2}\delta \phi_k/M_{\rm pl}$
and $\bar{\delta \chi}_k=k^{3/2}\delta \chi_k/M_{\rm pl}$
with the wave number $k=m$
for $m=1.16 \times 10^{-3} M_{\rm pl}$ and 
$M=1.39 \times 10^{-5} M_{\rm pl}$.
We choose the background initial conditions 
$\phi=0.1 M_{\rm pl}$, $\chi=1.0 \times 10^{-3} M_{\rm pl}$, 
$\dot{\phi}=-8.48 \times 10^{-4}mM_{\rm pl}$, and 
$\dot{\chi}=1.18 \times 10^{-5}mM_{\rm pl}$.
\label{fig5}}
\end{figure}

\begin{figure}
\includegraphics[height=3.0in,width=3.1in]{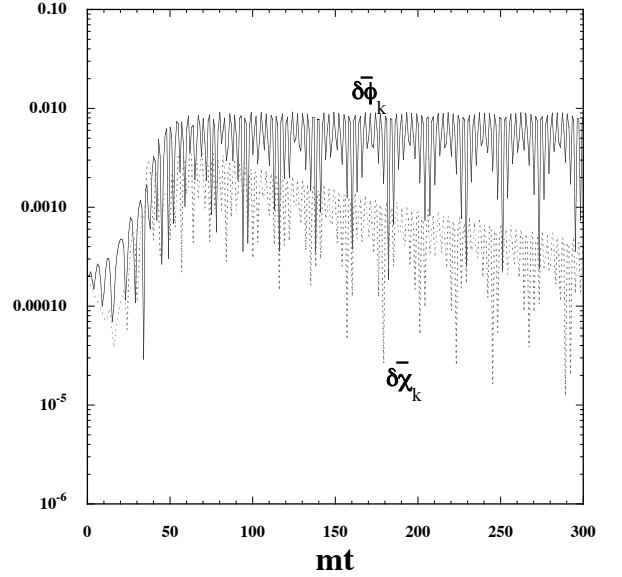}
\caption{
Evolution of the field perturbations
with the wave number $k=m$
for $m=2.89 \times 10^{-4}  M_{\rm pl}$ and 
$M=1.46 \times 10^{-5}M_{\rm pl}$.
We choose the background initial conditions 
$\phi=0.1M_{\rm pl}$, $\chi=1.0 \times 10^{-3}M_{\rm pl}$, 
$\dot{\phi}=-7.35 \times  10^{-3} mM_{\rm pl}$, and 
$\dot{\chi}=6.85 \times 10^{-4} mM_{\rm pl}$.
\label{fig6}}
\end{figure}

We introduce the rescaled fields $\delta \varphi_k=a^{3/2} \delta \phi_k$
and $\delta X_k=a^{3/2}e^{b} \delta \chi_k$ to estimate the growth of 
perturbations in the regime (\ref{phicon}).
Neglecting the contributions of the r.h.s. of Eqs.~(\ref{delphieq})
and (\ref{delchieq}) and also using the approximation 
$e^{-2b} \simeq 1$ in the regime $H \ll m$, the field perturbations 
$\delta \varphi_k$ and $\delta X_k$
obey the following equations
\ba
& &\frac{d^2}{dz^2} \delta \varphi_k+
\left[ A_k-2q_{\phi} \cos (2z) \right]  \delta \varphi_k \simeq 0\,,
\label{delvarphi} \\
& &\frac{d^2}{dz^2} \delta X_k+
\left[ A_k-2q_{\chi} \cos (2z) \right]  \delta X_k \simeq 0\,,
\label{delX} 
\ea
where $2z=mt+\pi/2$.
The quantities $A_k$, $q_{\phi}$, and $q_{\chi}$ are given by 
\ba
A_k &=& 4+4 \frac{k^2}{m^2 a^2}\,,\\
q_{\phi} &=& 
\frac{20\sqrt{6}}{7} 
\left(1-\frac{7M^2}{10m^2}
\right) \left( \frac{m}{M} \right)^2  
\frac{\phi_i}{M_{\rm pl}} \frac{\pi/2}{mt}\,,\\
q_{\chi} &=& \frac{20\sqrt{6}}{21} 
\left( \frac{m}{M} \right)^2 
\frac{\phi_i}{M_{\rm pl}} \frac{\pi/2}{mt}\,,
\ea
which are time-dependent.

Equations (\ref{delvarphi}) and (\ref{delX}) are the so-called
Mathieu equations describing parametric resonance caused by
the oscillation of the field $\phi$ \cite{Robert,KLS}
(see also the review \cite{Bassett}).
In the regime (\ref{phicon}) both $q_{\phi}$ and 
$q_{\chi}$ are smaller than 1 for $t \ge t_i=\pi/(2m)$.
In this case the resonance occurs in narrow bands 
near $A_k=l^2$, where $l=1, 2, \cdots$ \cite{KLS,Tsuji}.
As the physical momentum $k/a$ redshifts away, the field
perturbations approach the instability band at $A_k=4$.
Although $\delta \varphi_k$ and $\delta X_k$ can be amplified
for $A_k \simeq 4$ and $q_{\phi} \lesssim 1$, $q_{\chi} \lesssim 1$, 
this narrow parametric resonance is not efficient enough to 
lead to the growth of $\delta \phi_k$ and $\delta \chi_k$
against the Hubble friction \cite{KLS}.

If the initial field $\phi_i$ satisfies the condition $(m/M)^2|\phi_i|/M_{\rm pl} \gg 1$, 
the quantities $q_{\phi}$ and $q_{\chi}$ are much larger than 1 at the onset 
of reheating. This corresponds to the so-called broad 
resonance regime \cite{KLS}
in which the perturbations $\delta \phi_k$ and $\delta \chi_k$
can grow even against the Hubble friction.
We caution, however, that Eqs.~(\ref{delvarphi}) and (\ref{delX}) are 
no longer valid because the background solution (\ref{infosc})
is subject to change due to the effect of higher-order terms
in the potential (\ref{Vtotal}).
Still, the non-adiabatic particle production occurs 
around the potential minimum ($\phi=0$) \cite{KLS}.
In this region the dominant contribution to the potential 
is the quadratic term $m^2 \phi^2/2$.
Hence it is expected that preheating can 
be efficient for the values of $q_{\phi}$ and $q_{\chi}$ 
much larger than 1 at the onset of the field oscillations.

Numerically we solve the perturbations equations (\ref{delphieq}) 
and (\ref{delchieq}) together with the background equations
(\ref{Hubeq}), (\ref{phieqm}), and (\ref{ddotchi}) 
for the full potential (\ref{Vtotal}) without using the approximate 
expression (\ref{Vre}).
In Figs.~\ref{fig5} and \ref{fig6} we plot the evolution of 
the field perturbations $\delta \phi_k$ and $\delta \chi_k$
with the wave number $k=m$ for two different choices of 
the parameters $m$ and $M$ (which are constrained by 
the WMAP normalization in Fig.~\ref{fig1}).
The initial conditions of the perturbations are chosen to 
recover the vacuum state characterized by 
$\delta \varphi_k(t_i)=e^{-i \omega_{\phi} t_i}/\sqrt{2\omega_{\phi}}$ 
and 
$\delta X_k(t_i)=e^{-i \omega_{\chi} t_i}/\sqrt{2\omega_{\chi}}$.

Figure \ref{fig5} corresponds to the mass scales
$m=1.16 \times 10^{-3}M_{\rm pl}$ and 
$M=1.39 \times 10^{-5}M_{\rm pl}$, i.e., 
the ratio $m/M=83$.
The field value at the onset of oscillations is found 
to be $\phi_i = 5.0 \times 10^{-3} M_{\rm pl}$, in which case
$q_{\phi}(t_i)=244$ and $q_{\chi}(t_i)=81$.
Figure \ref{fig5} shows that both $\delta \phi_k$ and $\delta \chi_k$ 
grow rapidly by the broad parametric resonance.
The growth of the field perturbations ends when $q_{\phi}$
and $q_{\chi}$ drop below 1.

Figure \ref{fig6} corresponds to the ratio $m/M=20$, 
in which case $\phi_i = 1.5 \times 10^{-2} M_{\rm pl}$, 
$q_{\phi}(t_i)=41$, and $q_{\chi}(t_i)=4.6$.
Compared to the evolution in Fig.~\ref{fig5}, 
preheating is less efficient because of the smaller
values of $q_{\phi}(t_i)$ and $q_{\chi}(t_i)$.
The parameter to control the efficiency of preheating 
is the mass ratio $m/M$.
For larger $m/M$ the creation of particles tends to be 
more significant.
For the mass $m$ smaller than $10^{-4} M_{\rm pl}$
the field perturbations $\delta \phi_k$ and $\delta \chi_k$
hardly grow against the Hubble friction 
because they are not in the broad resonance regime.

In our numerical simulations we did not take into account 
the rescattering effect between different modes of the 
particles. The lattice simulation \cite{lattice,Felder} is required 
to deal with this problem.
It will be of interest to see how the non-linear effect can 
affect the evolution of perturbations 
at the final stage of preheating.

\section{Conclusions}
\label{concludesec}

We have studied the viability of the $f(R)$ inflationary scenario
in the context of $F({\cal R})$ supergravity.
In the high-energy regime characterized by the condition $|R| \gg R_0$ 
there is a correction of the form $(-R)^{3/2}/m$ 
to the function $f(R)=3R/10-R^2/(6M^2)$.
Introducing the dimensionless functions $\alpha$ and $\beta$
in Eqs.~(\ref{albe}), we showed that these are constrained to be
in the range (\ref{alcon}) to realize inflation with the number of 
e-foldings $N$.

The masses of the scalaron field in the regimes $|R| \gg R_0$
and $|R| \ll R_0$ are approximately given by $M$ and 
$m$, respectively. {}From the WMAP normalization of the 
CMB temperature anisotropies we derived 
$M$ and $m$ as a function of $\alpha$ in Fig.~\ref{fig1}.
The weak dependence of $M$ with respect to $\alpha$
means that the term $-R^2/(6M^2)$ needs to dominate
over the correction $(-R)^{3/2}/m$ during inflation.
We also showed that the model is within the 1$\sigma$ 
observational contour constrained from the joint data
analysis of WMAP7, BAO, and HST, 
by evaluating the scalar spectral index $n_{\rm s}$
and the tensor-to-scalar ratio $r$.

In the presence of the pseudo-scalar field $\chi$ coupled to 
the scalaron field $\phi$ we derived the effective potential (\ref{Vtotal}) 
and their kinetic energies (\ref{kine2}) 
in the low-energy regime ($|R| \ll R_0$).
Provided that the condition (\ref{phic}) is satisfied, the effective potential 
has two local minima at $\chi=\pm \chi_c$.
Around the global minimum at $\phi=\chi=0$ 
the system is described by two massive scalar fields
with other interaction terms given in Eq.~(\ref{Vre}).
Even if $\chi$ is initially close to 0, $\chi$ typically catches up 
to $\phi$ around the onset of the field oscillations (see Fig.~\ref{fig4}).

In the regime where the field $\phi$ is in the range (\ref{phicon})
we showed that both the field perturbations 
$\delta \varphi_k=a^{3/2} \delta \phi_k$ and 
$\delta X_k=a^{3/2}e^b \delta \chi_k$ obey the Mathieu 
equations (\ref{delvarphi}) and (\ref{delX}).
This corresponds to the narrow resonance regime in which 
$q_{\phi}$ and $q_{\chi}$ are smaller than the order of unity.
The broad resonance regime is characterized by the condition
$(m/M)^2 |\phi|/M_{\rm pl} \gg 1$, but in this case the expansion 
(\ref{Vre}) of the effective potential around the minimum is
no longer valid. In order to confirm the presence of 
the broad resonance we numerically solved the perturbation equations
(\ref{delphieq}) and (\ref{delchieq}) for the full potential (\ref{Vtotal}). 
Indeed we found that preheating of both the perturbations 
$\delta \phi_k$ and $\delta \chi_k$ is efficient in this regime.
As we see in Figs.~\ref{fig5} and \ref{fig6}, the broad parametric 
resonance is more significant for larger values of $m/M$.

Our results lend compelling support to the phenomenological 
viability of the bosonic sector of $F(\car)$ supergravity, 
in addition to its formal consistency. It is also worthwhile to
recall that supergravity unifies bosons and fermions with General Relativity,
highly constrains particle spectrum and interactions, has the ideal candidate
for a dark matter particle such as the lightest super-particle \cite{dm}, 
and can be deduced from quantum gravity such as superstring theory. The $F(\car)$ 
supergravity action (\ref{frsup}) is truly chiral in superspace, so that it is
expected to be protected against quantum corrections \cite{Kreview}, which is 
important for stabilizing the masses $M$ and $m$ in quantum theory. 

After the broad parametric resonance and the subsequent rescattering at the
final stage of preheating, decay of super-scalaron produces new particles 
including the visible sector. 
It can be studied perturbatively, when adding a supersymmetric 
matter action to the $F({\cal R})$ supergravity action in Eq.~(\ref{frsup}), 
along the standard lines, see e.g., Section 13 of Ref.~\cite{Kreview} 
for a review.  As is well known, the Starobinsky $f(R)$ gravity model has 
the universal reheating mechanism due to the coupling of scalaron 
to all matter \cite{Starobinsky80,S82}.
The same applies to super-scalaron in
matter coupled $F({\cal R})$ supergravity \cite{Ketov11}. 

Of course, our analysis of $F(\car)$ supergravity is still incomplete since it
does not include fermions such as gravitinos and inflatinos (gravitino is the
fermionic superpartner of graviton, and inflatino is the fermionic superpartner of scalaron). 
We did not study in this paper further particle
production to complete the reheating process. 
A calculation of decay rates and particle abundances after preheating 
requires an extension of the action (\ref{frsup}) by some hidden sector 
to be responsible for supersymmetry breaking. It should be 
the subject of a separate investigation.

\section*{ACKNOWLEDGEMENTS}
SVK is supported by the World Premier International 
Research Center Initiative (WPI Initiative), MEXT, Japan.
ST is supported by the Grant-in-Aid for Scientific Research Fund of the JSPS 
No.~30318802 and the Fund for Scientific Research on Innovative Areas 
(JSPS No.~21111006). 

\end{document}